\documentstyle[12pt]{article}

\def\photonatomup{\begin{picture}(1.5,3)(0,0)
                             \put(-0.75,3){\tencircw \symbol{3}}
                             \put(-0.75,1.5){\tencircw \symbol{2}}
                             \put(0.75,1.5){\tencircw \symbol{0}}
                             \put(0.75,0){\tencircw \symbol{1}}
                   \end{picture}
                  }

\def\photonuphalf{\begin{picture}(1.5,15)(0,0)
                      \multiput(0,0)(0,3){5}{\photonatomup}
                   \end{picture}
                  }

\title{
\begin{flushright}
{\normalsize Yaroslavl State University\\
             Preprint YARU-HE-99/01\\
             hep-ph/9903210} \\
\end{flushright}
Comment on the paper: ``Neutrino pair production by a virtual photon 
in an external magnetic field'' by Zhukovskii et al.}

\author{A.V.~Kuznetsov\thanks{avkuzn@uniyar.ac.ru}, 
N.V.~Mikheev\thanks{mikheev@yars.free.net}\\
{\small\it Division of Theoretical Physics, Department of Physics,}\\
{\small\it Yaroslavl State University, Sovietskaya 14,}\\
{\small\it 150000 Yaroslavl, Russian Federation.}}

\date{}

\begin{document}

\maketitle

\begin{abstract}
We point out some serious mistakes in the investigation of Zhukovskii et al. 
Both the amplitude and the probability of the process were calculated 
wrongly, that is, the problem of the neutrino pair production by a 
virtual photon in an external magnetic field is still unsolved. 
\end{abstract}

\newpage

In the recent paper~\cite{Zh97} an attempt was made of investigation of 
the virtual photon decay into the neutrino pair in an external magnetic 
field with a strength much smaller than the critical value, in the frame 
of the standard 
model with neutrino mixing. However, the results of the calculations were 
incorrect. The basic formula (1) for the process amplitude was written 
in a rather slipshod manner, namely, the summation over the lepton flavor 
$(a = e, \mu, \tau)$ was defined inadequately, besides the quark 
contribution into a part of the aplitude, which is diagonal with respect 
to a neutrino type (proportional to $\delta_{ij}$), was not taken into 
account. It is also seen from Eq.(1) that the authors extended 
incorrectly their result~\cite{Zh96} for the crossed process 
$\nu_i \to \nu_j \gamma \; (i \ne j)$, where the charged current contribution 
was presented only
\footnote{In our opinion, Eq.(4) of the paper~\cite{Zh96} for the matrix 
element contains an extra factor 2 which is repeated in the commented 
formula (1).}, 
on a case $i = j$, where the $Z$ exchange contribution 
was also presented. Really, the Eq.(1) demonstrates manifestly that the 
authors have ``discovered'' a new type of the effective local Lagrangian of 
the neutrino interaction with charged leptons of a chiral type (when only 
left-handed charged leptons interact). However, as was known up to 
now~\cite{RPP}, the Lagrangian of such an interaction had a form
\begin{eqnarray}
- {\cal L}^{(Z)}_{\nu e}  =  \frac{ G_F }{ \sqrt{2} } \;
[\bar \nu \gamma^{\alpha} ( 1 - \gamma_5 ) \nu] \;
[ \bar e \gamma_{\alpha} ( g^{\nu e}_V - g^{\nu e}_A \gamma_5 ) e],
\label{eq:L}
\end{eqnarray}

\noindent
where
$$ g^{\nu e}_V = - \frac{1}{2}  + 2 \sin^2\theta_W , \qquad
g^{\nu e}_A = - \frac{1}{2}. $$

\noindent
By this means both left-handed and right-handed charged leptons take part in 
the interaction $(g^{\nu e}_V \ne g^{\nu e}_A)$. 

Our next remark is concerned with a procedure of calculations with using of 
the effective local Lagrangian of weak interactions. 
The matter is, that taking of the local limit leads to an appearance of 
the two problems: an amplitude acquires the ultraviolet divergency 
in this limit, and the triangle axial-vector anomaly as well. 
It is most easily seen if an amplitude 
is expanded into a series in terms of the external magnetic
field as it is shown in~Fig.~1,
where the dashed lines designate the external field. 

\begin{figure}[htb]
\unitlength 1mm
\linethickness{0.4pt}
\begin{picture}(133.00,60.00)
\linethickness{0.8pt}
\put(20.00,25.00){\circle{10.00}}
\put(20.00,25.00){\circle{10.05}}
\put(20.00,25.00){\circle{10.10}}
\put(20.00,25.00){\circle{10.15}}
\put(20.00,25.00){\circle{13.00}}
\put(20.00,25.00){\circle{13.05}}
\put(20.00,25.00){\circle{13.10}}
\put(20.00,25.00){\circle{13.15}}
\linethickness{0.4pt}
\put(20.00,18.50){\line(+1,-2){7}}
\put(20.00,18.50){\line(-1,-2){7}}
\put(20.00,31.50){\photonuphalf}
\put(20.00,18.50){\circle*{1.0}}
\put(20.00,31.50){\circle*{1.0}}
\put(24.00,45.00){\makebox(0,0)[cc]{\large $\gamma$}}
\put(17.00,7.00){\makebox(0,0)[cc]{\large $\nu$}}
\put(23.00,7.00){\makebox(0,0)[cc]{\large $\nu$}}
\put(35.00,25.00){\makebox(0,0)[cc]{\LARGE =}}
\put(50.00,25.00){\circle{13.00}}
\put(50.00,18.50){\line(+1,-2){7}}
\put(50.00,18.50){\line(-1,-2){7}}
\put(50.00,31.50){\photonuphalf}
\put(50.00,18.50){\circle*{1.0}}
\put(50.00,31.50){\circle*{1.0}}
\put(54.00,45.00){\makebox(0,0)[cc]{\large $\gamma$}}
\put(47.00,7.00){\makebox(0,0)[cc]{\large $\nu$}}
\put(53.00,7.00){\makebox(0,0)[cc]{\large $\nu$}}
\put(65.00,25.00){\makebox(0,0)[cc]{\LARGE +}}
\put(80.00,25.00){\circle{13.00}}
\put(80.00,18.50){\line(+1,-2){7}}
\put(80.00,18.50){\line(-1,-2){7}}
\put(80.00,31.50){\photonuphalf}
\put(80.00,18.50){\circle*{1.0}}
\put(80.00,31.50){\circle*{1.0}}
\multiput(76.70,30.30)(-3.00,+6.00){3}{\line(-1,+2){2}}
\put(76.70,30.30){\circle*{1.0}}
\put(68.70,46.30){\makebox(0,0)[cc]{\Large $\times$}}
\put(84.00,45.00){\makebox(0,0)[cc]{\large $\gamma$}}
\put(75.00,48.00){\makebox(0,0)[cc]{\large $A^{ext}$}}
\put(77.00,7.00){\makebox(0,0)[cc]{\large $\nu$}}
\put(83.00,7.00){\makebox(0,0)[cc]{\large $\nu$}}
\put(95.00,25.00){\makebox(0,0)[cc]{\LARGE +}}
\put(110.00,25.00){\circle{13.00}}
\put(110.00,18.50){\line(+1,-2){7}}
\put(110.00,18.50){\line(-1,-2){7}}
\put(110.00,31.50){\photonuphalf}
\put(110.00,18.50){\circle*{1.0}}
\put(110.00,31.50){\circle*{1.0}}
\multiput(106.70,30.30)(-3.00,+6.00){3}{\line(-1,+2){2}}
\multiput(113.30,30.30)(+3.00,+6.00){3}{\line(+1,+2){2}}
\put(106.70,30.30){\circle*{1.0}}
\put(113.30,30.30){\circle*{1.0}}
\put(121.30,46.30){\makebox(0,0)[cc]{\Large $\times$}}
\put(98.70,46.30){\makebox(0,0)[cc]{\Large $\times$}}
\put(114.00,45.00){\makebox(0,0)[cc]{\large $\gamma$}}
\put(105.00,48.00){\makebox(0,0)[cc]{\large $A^{ext}$}}
\put(128.00,48.00){\makebox(0,0)[cc]{\large $A^{ext}$}}
\put(107.00,7.00){\makebox(0,0)[cc]{\large $\nu$}}
\put(113.00,7.00){\makebox(0,0)[cc]{\large $\nu$}}
\put(125.00,25.00){\makebox(0,0)[cc]{\LARGE +}}
\put(130.00,25.00){\makebox(0,0)[lc]{\LARGE $\cdots$}}
\end{picture}
\caption{}
\end{figure}

The zero term in this expansion 
\begin{eqnarray}
{\cal M}^{(0)} = {\cal M} (B = 0), 
\label{eq:M0}
\end{eqnarray}

\noindent 
contains the ultraviolet divergency, while the term, linear on the external 
field
\begin{eqnarray}
{\cal M}^{(1)} = B \frac{d{\cal M}}{dB} 
\bigg \vert_{ B = 0 } 
\label{eq:M1}
\end{eqnarray}

\noindent 
contains the triangle anomaly because of the presence of the 
axial-vector coupling in the effective weak Lagrangian.
To obtain a correct expression for the amplitude one should perform a 
procedure of the two-step subtraction
\begin{eqnarray}
{\cal M}_{corr} = \left( {\cal M} - {\cal M}^{(0)} - {\cal M}^{(1)} \right) 
+ \widetilde {\cal M}^{(0)} + \widetilde {\cal M}^{(1)},
\label{eq:Mcor}
\end{eqnarray}

\noindent 
where the correct zero-field term $\widetilde {\cal M}^{(0)}$ and the term 
$\widetilde {\cal M}^{(1)}$ linear on the field should be found 
independently without taking the local limit and with taking account of 
the neutrino interaction via $W$ and $Z$ bosons with all charged fermions, 
both leptons and quarks. 
For example, the expression for $\widetilde {\cal M}^{(1)}$ can be obtained 
from the amplitude of the Compton-like process 
$\nu + \gamma^* \to \nu + \gamma^*$ \cite{KM} where the field tensor of one 
photon is replaced by the external field tensor. 
The subtraction procedure of this kind is performed in our paper~\cite{VKM}. 
Since the authors of the commented paper do not take the triangle anomaly 
problem into account, their result would be incorrect even if the proper 
Lagrangian of the lepton-neutrino interaction was used. 

Finally, the statement of the authors \cite{Zh97} about the applicability 
of their Eqs.(3), (4) for an analysis of a real photon decay into the neutrino 
pair is radically incorrect. It could be seen from the kinematical arguments. 
Really, these formulas were written for a photon with the space-like 
momentum, while the total momentum of a neutrino pair is always the 
time-like one. In such a case this process is kinematically forbidden 
not only in vacuum but in the constant external magnetic field as well. 

To summarize: In our opinion, the problem of the neutrino pair production by 
a virtual photon in an external magnetic field is still unsolved.

\end{document}